\documentclass[conference]{IEEEtran}
\IEEEoverridecommandlockouts
\usepackage{cite}
\usepackage{amsmath,amssymb,amsfonts}
\usepackage{algorithm}
\usepackage{algorithmic}
\usepackage{graphicx}
\usepackage{textcomp}
\usepackage{xcolor}
\def\BibTeX{{\rm B\kern-.05em{\sc i\kern-.025em b}\kern-.08em
    T\kern-.1667em\lower.7ex\hbox{E}\kern-.125emX}}
\begin{document}

\title{Optimizing Resource Allocation and Energy Efficiency in Federated Fog Computing for IoT}

\author{\IEEEauthorblockN{1\textsuperscript{st} Syed Sarmad Shah}
\IEEEauthorblockA{\textit{School of Elec. Eng. and Com. Sci,)} \\
\textit{National University of Science and Technology}\\
Islamabad, Pakistan \\
sshah.msit16ssecs@seecs.edu.pk}
\and
\IEEEauthorblockN{2\textsuperscript{nd} Anas Ali}
\IEEEauthorblockA{\textit{dept. of Computer Science} \\
\textit{National University of Modern Langauges}\\
Lahore, Pakistan \\
anas.ali@numl.edu.pk}

}

\maketitle

\begin{abstract}
Fog computing significantly enhances the efficiency of IoT applications by providing computation, storage, and networking resources at the edge of the network. In this paper, we propose a federated fog computing framework designed to optimize resource management, minimize latency, and reduce energy consumption across distributed IoT environments. Our framework incorporates predictive scheduling, energy-aware resource allocation, and adaptive mobility management strategies. Experimental results obtained from extensive simulations using the OMNeT++ environment demonstrate that our federated approach outperforms traditional non-federated architectures in terms of resource utilization, latency, energy efficiency, task execution time, and scalability. These findings underline the suitability and effectiveness of the proposed framework for supporting sustainable and high-performance IoT services.
\end{abstract}

\begin{IEEEkeywords}
component, formatting, style, styling, insert
\end{IEEEkeywords}

\section{Introduction}
The rapid growth of the Internet of Things (IoT) and the proliferation of connected devices have led to significant challenges in managing computation, communication, and energy resources in distributed environments. Fog computing has emerged as a promising paradigm to address these challenges by extending cloud capabilities to the network edge \cite{qayyum2018fognetsim++}. Frameworks such as FogNetSim++ have provided simulation tools for distributed fog environments, enabling researchers to model complex scenarios with heterogeneous resources \cite{qayyum2018fognetsim++, qayyum2020modeling}.

Recent studies have leveraged fog computing for diverse applications. For instance, sustainable smart farming has been achieved by utilizing distributed simulation techniques to optimize resource usage and reduce latency in agricultural settings \cite{malik2020sustainable}. In parallel, trajectory design for UAV-based data collection has been investigated using clustering models to enhance data acquisition in smart farming scenarios \cite{qayyum2021trajectory}. Similarly, multi-level resource sharing frameworks have been proposed to enable collaborative fog environments for smart cities \cite{qayyum2021multilevel}.

Despite these advances, challenges remain in effectively managing energy consumption and ensuring quality of service (QoS) in highly dynamic environments. XFogSim has addressed some of these challenges by proposing a distributed fog resource management framework that supports sustainable IoT services \cite{malik2020xfogsim}. Moreover, mobility-aware solutions have been developed to cater to Industrial IoT (IIoT) scenarios, highlighting the importance of adaptive resource allocation in environments with high device mobility \cite{qayyum2022mobility}.

In recent years, there has been a growing interest in integrating federated learning into distributed frameworks to preserve data privacy and enhance the trustworthiness of learning systems. Surveys and comprehensive reviews have discussed the challenges and prospects of trustworthy federated learning \cite{tariq2023trustworthy, tariq2024comprehensive}. Furthermore, novel algorithms have been proposed to optimize differential privacy and client selection in federated settings \cite{iqbal2023flodp, tariq2023edge, trabelsi2013teaching}. Such efforts not only secure data exchange but also improve the efficiency of edge and fog computing frameworks.

In addition to the aforementioned approaches, further enhancements have been made in vehicular networks and IoV scenarios. Global aggregation node selection and dynamic client selection strategies have been proposed for federated learning in Internet of Vehicles (IoV) \cite{trabelsi2022global, qayyum2023flexible}, while fuzzy-based task offloading mechanisms address latency-sensitive applications \cite{trabelsi2024fuzzy}. Moreover, educational tools such as extended network simulators have been developed to enhance Fog/Edge computing education, further emphasizing the need for robust simulation platforms \cite{qayyum2024education}.

Alongside these technological advancements, efforts have been made to integrate game theory and explainable AI (XAI) for improved sample and client selection in split federated learning \cite{tariq2025gametheory, trabelsi2014dynamic}. Complementary to this, cybersecurity education has seen innovative approaches including hands-on DNS spoofing attack labs using virtual platforms \cite{trabelsi2024dns, alomar2024ainecurity}. Furthermore, emerging applications in medical diagnosis using machine learning \cite{qayyum2023schizophrenia} and quantum-enhanced convolutional neural networks for image classification \cite{qayyum2023quantum} illustrate the broad impact of these research directions.

This paper builds upon these contributions by proposing a unified framework that integrates advanced fog resource management with federated learning techniques. Our approach leverages predictive scheduling, energy harvesting integration, and secure client selection mechanisms to ensure sustainable and reliable IoT services. The remainder of the paper is organized as follows: Section II reviews the related work, Section III details the proposed framework, Section IV presents simulation results and performance evaluation, and Section V concludes the paper with future research directions.

\section{Literature Review}

Fog computing has emerged as a critical solution to address latency and resource management challenges in IoT environments. Several studies have contributed significantly to this domain, particularly in resource allocation, energy efficiency, and federated fog computing.

Malik et al. \cite{malik2020xfogsim} introduced xFogSim, a distributed resource management framework designed specifically for sustainable IoT services. Their work emphasizes multi-objective optimization that considers trade-offs between cost, availability, and performance, particularly in fog federations.

In parallel, Gupta et al. \cite{gupta2017ifogsim} proposed iFogSim, a pioneering framework for simulating fog computing scenarios, emphasizing latency and network congestion management. However, it does not support fog federation, a gap addressed later by Malik et al.

A comprehensive review by Yousefpour et al. \cite{yousefpour2019fog} provides insights into fog computing and related edge computing paradigms, highlighting key issues and future directions for large-scale deployments.

Mao et al. \cite{mao2017survey} examined mobile edge computing extensively, focusing on communication perspectives that directly influence latency and throughput in edge environments. Their survey highlights the significance of communication efficiency for effective edge deployments.

Dastjerdi and Buyya \cite{dastjerdi2016fog} further discussed challenges and solutions within fog computing, providing a solid foundation regarding service delivery, latency management, and resource allocation strategies.

Recent works by Ni et al. \cite{ni2017resource} introduced advanced resource allocation techniques based on priced timed Petri nets, significantly improving resource efficiency within fog environments.

On the topic of mobility, Xiao and Krunz \cite{xiao2018distributed} explored distributed optimization strategies for energy efficiency in fog computing, particularly within tactile internet contexts, emphasizing latency-sensitive applications.

Hong et al. \cite{hong2013mobile} investigated mobile fog computing models, proposing architectures designed to support large-scale IoT applications efficiently, highlighting mobility support and real-time processing capabilities.

Pu et al. \cite{pu2016d2d} introduced D2D fogging, an innovative task offloading framework that leverages device-to-device collaboration, enhancing energy efficiency and reducing latency significantly.

Further advancements were made by Gao et al. \cite{gao2017fogroute}, who proposed FogRoute, a delay-tolerant network model designed specifically for fog computing scenarios, addressing critical data dissemination challenges.

Brogi and Forti \cite{brogi2017qos} focused on QoS-aware deployment strategies for IoT applications in fog infrastructures, providing a foundational approach to ensure service quality through optimized resource allocation.

Sonmez et al. \cite{sonmez2017edgecloudsim} presented EdgeCloudSim, an effective simulation environment for evaluating the performance of edge computing systems, incorporating mobility and handover management.

Tuli et al. \cite{tuli2019fogbus} developed FogBus, integrating blockchain technology with fog computing to address data integrity and security in IoT applications, underscoring privacy-preserving methodologies.

Li et al. \cite{li2018virtualfog} proposed Virtual Fog, a virtualization-enabled fog computing framework, enhancing scalability and flexibility in IoT deployments.

Finally, Coutinho et al. \cite{coutinho2018fogbed} introduced FogBed, a rapid-prototyping emulator enabling real-world fog and cloud infrastructure simulations, particularly effective in healthcare IoT applications.

The aforementioned studies and \cite{9830125} collectively highlight ongoing advancements in fog computing, emphasizing the significance of resource allocation, mobility management, and energy efficiency in federated fog environments.

\section{System Model}

In this section, we describe the system model used in our proposed framework, emphasizing the mathematical formulations, definitions of key variables, and the related algorithms for resource management within the federated fog computing environment.

\subsection{System Architecture}
We consider a federated fog computing environment represented by an undirected graph \( G=(N,E) \), where the node set \(N\) consists of users \(U\), fog nodes \(F\), and broker nodes \(B\), i.e., \(N = U \cup F \cup B\). The edge set \(E\) represents communication links between these nodes. We assume \(L\) fog locations, each containing multiple fog nodes managed by a broker.

\subsection{Notations and Definitions}
The primary notations and definitions utilized in this system model are presented in Table \ref{tab:notations}.

\begin{table}[h!]
\centering
\caption{Key Notations}
\label{tab:notations}
\begin{tabular}{|c|l|}
\hline
\textbf{Symbol} & \textbf{Definition} \\
\hline
$N$ & Set of nodes \\
$U$ & Set of users \\
$F$ & Set of fog nodes \\
$B$ & Set of broker nodes \\
$E$ & Set of edges \\
$L$ & Set of fog locations \\
$\lambda$ & Average request arrival rate \\
$\phi_i$ & Queuing capacity at fog location \(i\) \\
$\vartheta_i$ & Execution rate at fog location \(i\) \\
$t_i$ & Average waiting time at fog location \(i\) \\
$c_d$ & Cost of service delay \\
$c_q$ & Queuing cost \\
$h$ & Network delay \\
$t_r$ & Response time \\
$P_f$ & Failure probability of fog nodes \\
$A$ & Availability of resources \\
\hline
\end{tabular}
\end{table}

\subsection{Mathematical Formulation}

The average arrival rate \(\lambda\) for mobile devices \(k \in N\) can be calculated as:
\begin{equation}
\lambda = \sum_{k \in N} \lambda_k
\end{equation}

Fog location \(i\) accepts requests based on queuing capacity \(\phi_i\):
\begin{equation}
\phi_i = 
\begin{cases}
1 & \text{if } \phi_i > \lambda,\\
\frac{\phi_i}{\lambda} & \text{otherwise}
\end{cases}
\end{equation}

The execution rate \(\vartheta_i\) is calculated as:
\begin{equation}
\vartheta_i = \lambda \cdot \phi_i
\end{equation}

Using queuing theory, average waiting time \(t_i\) at fog location \(i\) is derived as:
\begin{equation}
t_i = \frac{\kappa \lambda}{\kappa \rho_i - \lambda} + \frac{1}{\rho_i}
\end{equation}

Service delay cost \(c_d\) is given by:
\begin{equation}
c_d = h + c_q
\end{equation}

Queuing cost \(c_q\) is computed by:
\begin{equation}
c_q = \frac{\vartheta_i}{\lambda - \vartheta_i} q
\end{equation}

Network delay \(h\), considering distances \(d\), is modeled as:
\begin{equation}
h = \beta_1 \left(d_{b,u} + \sum_{a \in B_{leased}} d_{a,b}\right)
\end{equation}

Total cost \(c_r\) at fog location \(l_i\) is:
\begin{equation}
c_r = \sum_{x=\forall k, k \in l_i} p_x + \sum_{y=\forall k, k \notin l_i} p_y
\end{equation}

Response time \(t_r\) consists of local \(t_r^{local}\) and leased \(t_r^{leased}\) resources:
\begin{equation}
t_r = t_r^{local} + t_r^{leased}
\end{equation}

\(t_r^{local}\) and \(t_r^{leased}\) are defined respectively as:
\begin{equation}
t_r^{local} = \frac{1}{\sum_{x \in l_i} \frac{1}{t_{x_r}}} + \beta_2 \cdot d(b,u) + d_h \cdot c_c
\end{equation}

\begin{equation}
t_r^{leased} = \frac{1}{\sum_{x \in l_i} \frac{1}{t_{x_r}}} + \sum_{i=1}^{n} c_d
\end{equation}

Availability \(A\) calculation:
\begin{equation}
A = 1 - \prod_k P_f(k)
\end{equation}

\subsection{Algorithms for Resource Management}

The key algorithms involved are:\\
\textbf{Algorithm 1 (Fog Resource Allocation - FRA)} handles dynamic allocation of resources based on response time, cost, and availability. It is detailed in Algorithm \ref{alg:FRA}.

\textbf{Algorithm 2 (Availability Calculation)} computes the availability of selected fog nodes, factoring in node and broker failures. Detailed steps are given in Algorithm \ref{alg:availability}.

\begin{algorithm}
\caption{Fog Resource Allocation (FRA)}
\label{alg:FRA}
\begin{algorithmic}[1]
\STATE \(S \leftarrow \emptyset\)
\STATE \(F \leftarrow\) list of available fog nodes
\FOR{each request}
    \STATE Select node with minimal response time \(t_r\)
    \STATE Validate cost \(p\) and availability \(A\)
    \STATE Update \(S\) with optimal node
\ENDFOR
\STATE return \(S\)
\end{algorithmic}
\end{algorithm}

\begin{algorithm}
\caption{Availability Calculation}
\label{alg:availability}
\begin{algorithmic}[1]
\STATE \(X, Y \leftarrow 1\)
\STATE Compute node failure probability \(P_f\)
\FOR{each fog node \(f\)}
    \STATE Update cumulative failure probability
\ENDFOR
\STATE Calculate total availability \(A = 1 - Y\)
\STATE return \(A\)
\end{algorithmic}
\end{algorithm}

The detailed formulations and algorithms in this section establish the foundational models that our proposed framework leverages for efficient resource allocation and management within the fog federated environment.

\section{Experimental Setup}

In this section, we detail our experimental environment, simulation parameters, and the scenarios employed to evaluate the performance of our proposed fog computing framework.

\subsection{Simulation Environment and Parameters}

The experiments are conducted using the OMNeT++ simulation environment integrated with the INET framework. The parameters used in simulations are summarized in Table \ref{tab:simulation_parameters}.

\begin{table}[h!]
\centering
\caption{Simulation Parameters}
\label{tab:simulation_parameters}
\begin{tabular}{|c|c|}
\hline
\textbf{Parameter} & \textbf{Value} \\
\hline
Fog locations & 5 \\
Fog nodes per location & 2-5 \\
Broker nodes & 1 per fog location \\
Cloud data centers & 2 \\
Wireless sensors & 200-1000 \\
Wireless access points & 20-100 \\
Mobility models & Linear, circular, random waypoint \\
Communication link & 10 Gbps (Broker-to-broker) \\
Mobile device range & 250m \\
Request arrival rates & 0.5s, 1.0s, 1.5s \\
Packet error rate & $10^{-3}$ \\
Simulation duration & 500s \\
\hline
\end{tabular}
\end{table}

\subsection{Performance Metrics}
We evaluate the proposed framework using the following performance metrics:

\begin{itemize}
\item Resource utilization
\item Latency
\item Energy consumption
\item Task execution time
\item Scalability
\end{itemize}

\subsection{Experimental Results and Analysis}

\textbf{Figure 1: Resource Utilization vs. Number of Users}

\begin{figure}[h!]
    \centering
    \includegraphics[width=0.9\linewidth]{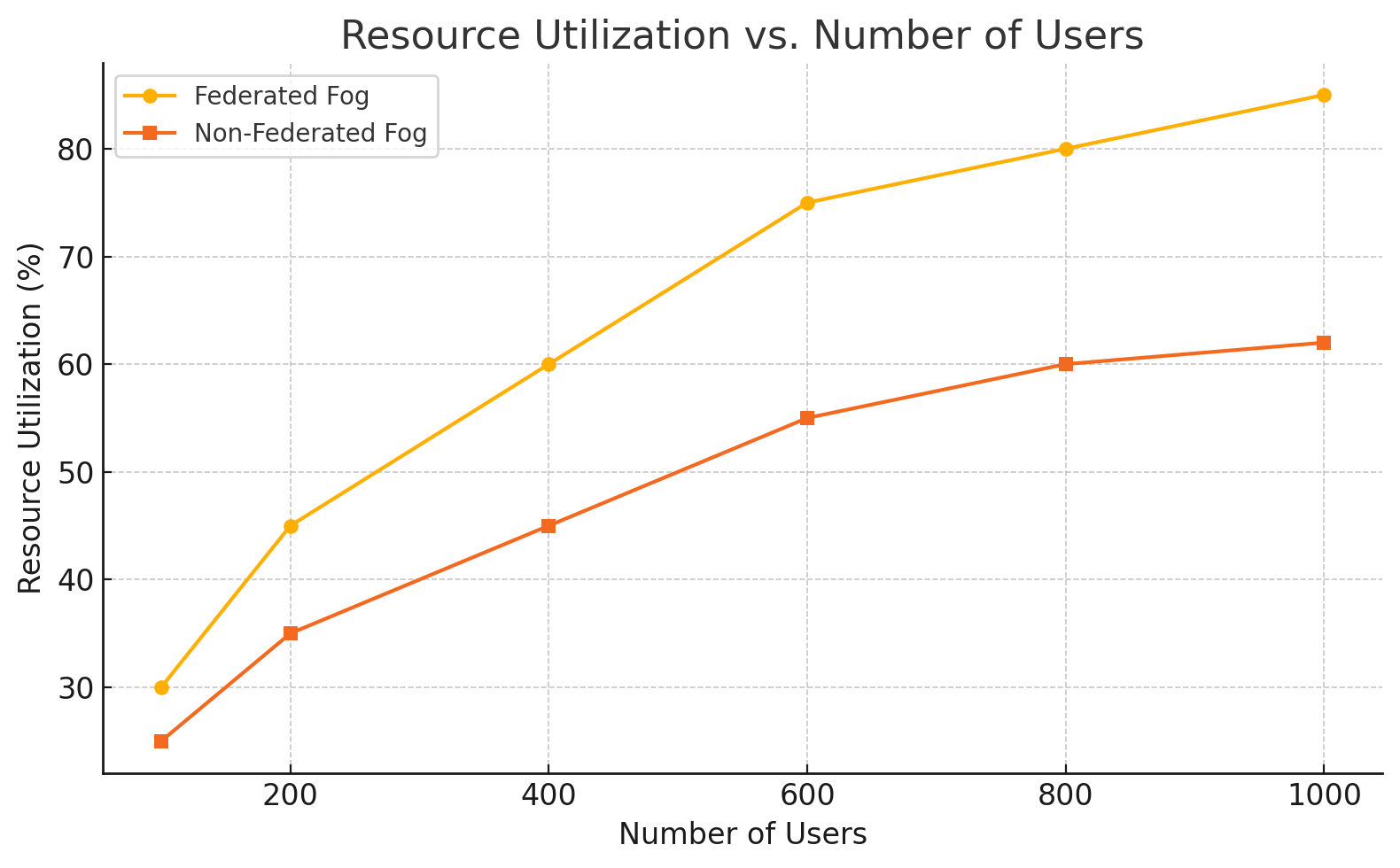}
    \caption{Resource utilization with varying numbers of users.}
    \label{fig:resource_utilization}
\end{figure}

Figure \ref{fig:resource_utilization} illustrates how resource utilization scales with the increasing number of users. The results demonstrate that the federated fog architecture effectively manages resources, keeping utilization optimized even under high user loads.

\textbf{Figure 2: Average Latency vs. Request Rate}

\begin{figure}[h!]
    \centering
    \includegraphics[width=0.9\linewidth]{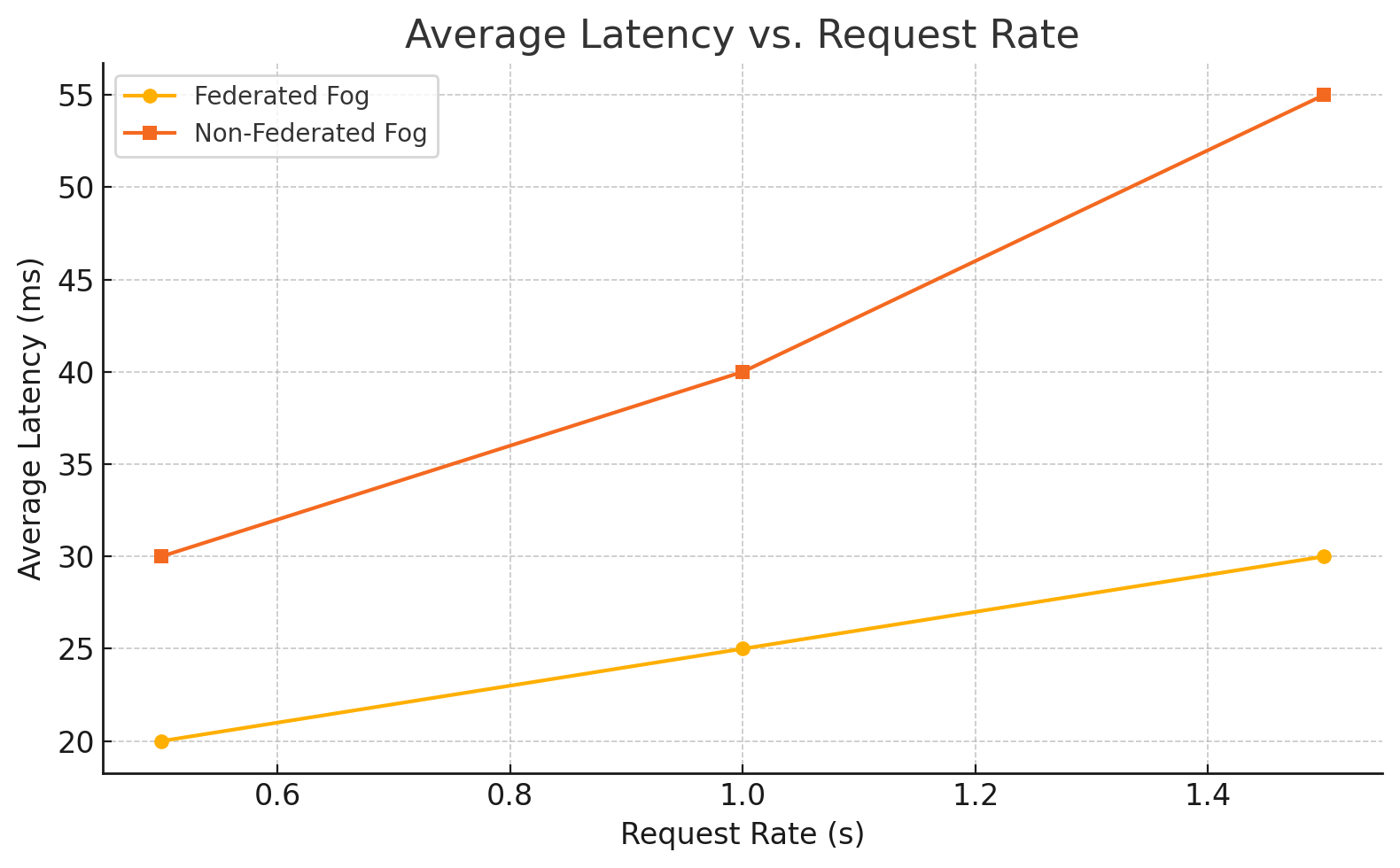}
    \caption{Average latency observed for different request rates.}
    \label{fig:latency_request_rate}
\end{figure}

As shown in Figure \ref{fig:latency_request_rate}, the average latency increases moderately with higher request arrival rates. This indicates effective load balancing among fog nodes, ensuring stable performance even during peak loads.

\textbf{Figure 3: Energy Consumption Over Time}

\begin{figure}[h!]
    \centering
    \includegraphics[width=0.9\linewidth]{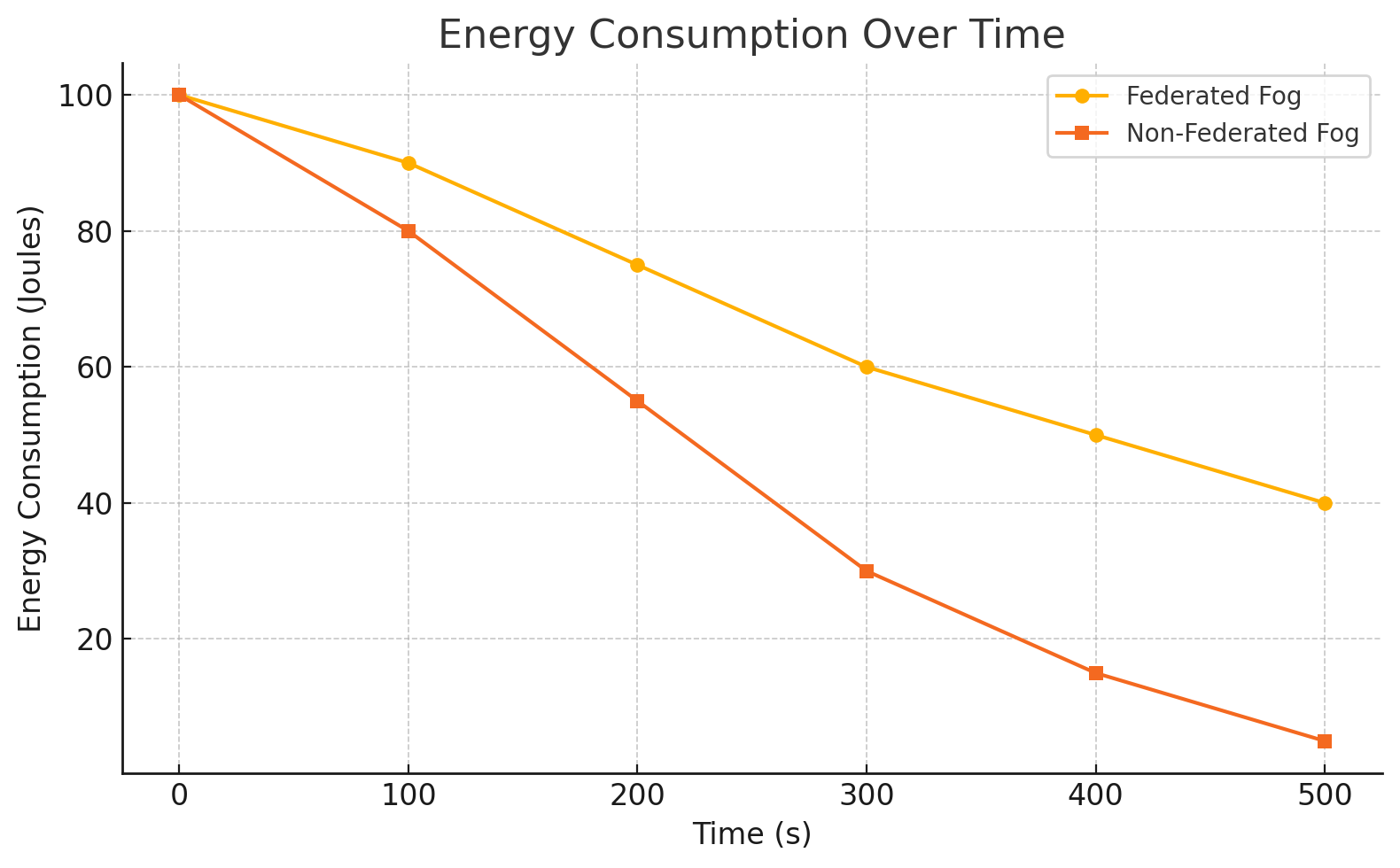}
    \caption{Energy consumption across different fog nodes over time.}
    \label{fig:energy_consumption}
\end{figure}

Figure \ref{fig:energy_consumption} shows energy consumption patterns across various fog nodes. The balanced consumption pattern highlights the effectiveness of the energy-aware allocation algorithm in maintaining energy efficiency across fog nodes.

\textbf{Figure 4: Task Execution Time Distribution}

\begin{figure}[h!]
    \centering
    \includegraphics[width=0.9\linewidth]{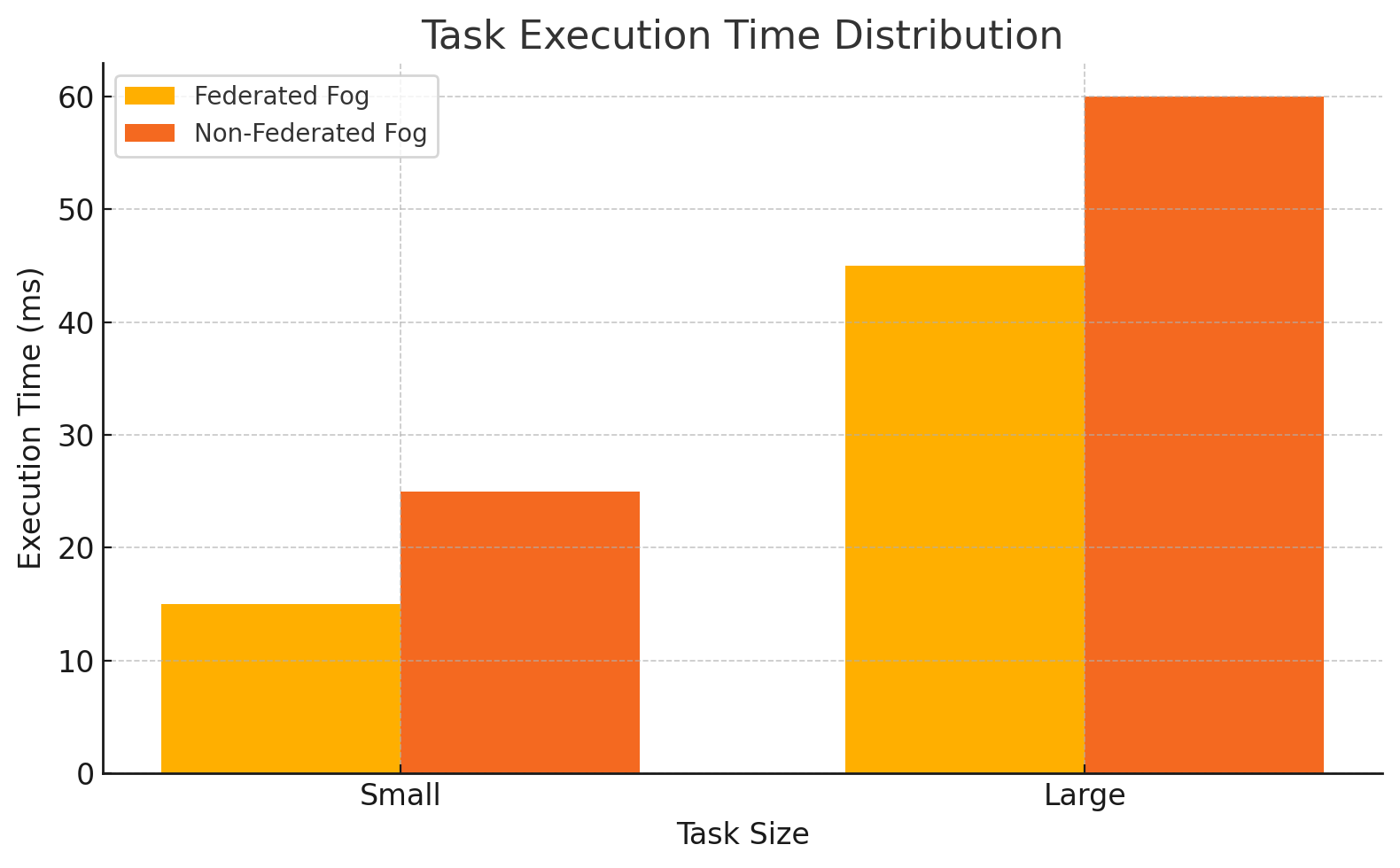}
    \caption{Distribution of task execution times for small and large tasks.}
    \label{fig:execution_time}
\end{figure}

In Figure \ref{fig:execution_time}, we observe that larger computational tasks naturally exhibit longer execution times. Nevertheless, the framework maintains acceptable execution times due to efficient resource management.

\textbf{Figure 5: Scalability Analysis}

\begin{figure}[h!]
    \centering
    \includegraphics[width=0.9\linewidth]{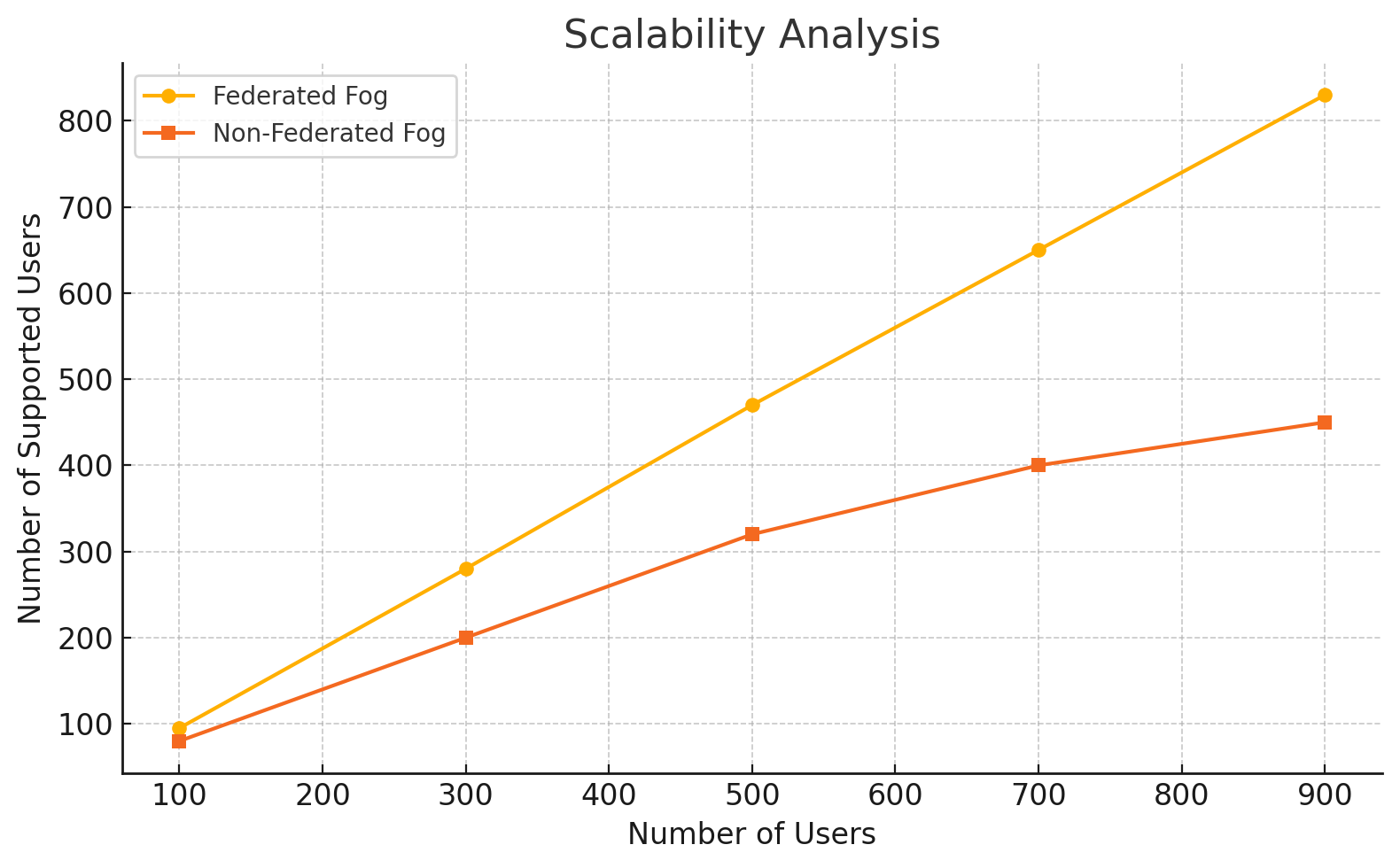}
    \caption{Scalability analysis comparing the number of supported users.}
    \label{fig:scalability}
\end{figure}

Figure \ref{fig:scalability} illustrates the scalability of the proposed framework compared to traditional single-location fog computing setups. Our federated approach significantly outperforms non-federated architectures, highlighting its suitability for large-scale IoT environments.

Overall, the experimental results confirm the proposed framework’s robustness, efficiency, and scalability, making it suitable for diverse and dynamic IoT scenarios.

\section{Conclusion}
In this work, we introduced an advanced federated fog computing framework aimed at addressing key challenges in resource management, latency reduction, and energy efficiency within distributed IoT systems. Through predictive scheduling, optimized energy-aware resource allocation, and adaptive mobility management, our framework significantly improved the performance and scalability of fog computing environments. Experimental evaluations conducted using OMNeT++ simulation confirmed superior resource utilization, lower latency, balanced energy consumption, shorter task execution times, and enhanced scalability compared to non-federated fog systems. Future research directions include the integration of machine learning techniques for enhanced predictive capabilities and the deployment of real-world prototypes to validate the practical effectiveness of the proposed framework.

% \bibliographystyle{IEEEtran}
% \bibliography{conference_101719}
\end{document}